\documentclass[pre,aps,twocolumn]{revtex4}

\usepackage{graphicx}
\usepackage{amsmath}
\usepackage{amssymb}
\usepackage{ifthen}
\usepackage{booktabs}
\usepackage{color}

\usepackage{graphicx}
\usepackage{dcolumn}
\usepackage{bm}
\usepackage{longtable}
\usepackage{gensymb}

\newcommand{\bb}{\begin{equation}}
\newcommand{\ee}{\end{equation}}
\newcommand{\ba}{\begin{eqnarray*}}
\newcommand{\ea}{\end{eqnarray*}}
\newcommand{\rhor}{\rho({\bf r})}
\newcommand{\dd}{{\rm d}}
\newcommand{\rr}{{\mathbf r}}
\newcommand{\dr}{{\rm d}{\bf r}}

\begin{document}

\title{Scaling of wetting and pre-wetting transitions on nano-patterned walls}

\author{Martin \surname{Posp\'\i\v sil}}
\affiliation{{Department of Physical Chemistry, University of Chemical Technology Prague, Praha 6, 166 28, Czech Republic;}\\
 {Department of Molecular and Mesoscopic Modelling, ICPF of the Czech Academy Sciences, Prague, Czech Republic}}

 \author{Martin \surname{L\'aska}}
\affiliation{{Department of Physical Chemistry, University of Chemical Technology Prague, Praha 6, 166 28, Czech Republic;}\\
 {Department of Molecular and Mesoscopic Modelling, ICPF of the Czech Academy Sciences, Prague, Czech Republic}}

\author{Andrew O. \surname{Parry}}
\affiliation{Department of Mathematics, Imperial College London, London SW7 2BZ, UK}

\author{Alexandr \surname{Malijevsk\'y}\footnote{malijevsky@icpf.cas.cz}}
\affiliation{
{Department of Physical Chemistry, University of Chemical Technology Prague, Praha 6, 166 28, Czech Republic;}\\
 {Department of Molecular and Mesoscopic Modelling, ICPF of the Czech Academy Sciences, Prague, Czech Republic}}

\begin{abstract}
\noindent We consider a nano-patterned planar wall consisting of a periodic array of stripes of width $L$, which are completely wet by liquid
(contact angle $\theta=0$), separated by regions of width $D$ which are completely dry (contact angle $\theta=\pi)$. Using microscopic Density
Functional Theory we show that in the presence of long-ranged dispersion forces, the wall-gas interface undergoes a first-order wetting transition,
at bulk coexistence, as the separation $D$ is reduced to a value $D_w\propto\ln L$, induced by the bridging between neighboring liquid droplets.
Associated with this is a line of pre-wetting transitions occurring off coexistence. By varying the stripe width $L$ we show that the pre-wetting
line shows universal scaling behaviour and data collapse. This verifies predictions based on mesoscopic models for the scaling properties associated
with finite-size effects at complete wetting including the logarithmic singular contribution to the surface free-energy.
\end{abstract}

\maketitle

\section{Introduction: wetting and bridging transitions}
As first discussed by Cahn \cite{cahn} and Ebner and Saam \cite{ebner} a wetting transition refers to the vanishing of the contact angle, $\theta$,
of a liquid drop on a substrate (wall), say, as the temperature $T$ is increased to a wetting temperature $T_w$ -- for reviews see for example
\cite{dietrich, schick, bonn}. Equivalently, and more microscopically, the transition refers to the divergence of the thickness of the adsorbed layer
of liquid at a planar wall-gas interface as $T$ approaches $T_w$ at bulk liquid-gas coexistence (chemical potential $\mu=\mu_{\rm sat}^-$). This can
also be thought of as the unbinding of the liquid-gas interface from the wall \cite{forgacs}. Wetting transitions can be continuous or first-order
depending on the sensitive balance and competition between the strength and range of the wall-fluid and fluid-fluid intermolecular forces. In most
situations wetting transitions are first-order in which case the thickness of the adsorbed liquid layer jumps from a microscopic to macroscopic value
at $T_w$; that is the liquid-gas interface unbinds discontinuously at $T_w$. Associated with this is a line of pre-wetting corresponding to
first-order transitions between thin and thick adsorbed liquid-like layers when the bulk gas is under-saturated $\mu<\mu_{\rm sat}$. In the
$T-\delta\mu$ plane, where $\delta\mu=\mu_{\rm sat}-\mu$, the line of pre-wetting transitions extends away tangentially from $T_w$ and terminates at
a pre-wetting critical point when the thin and thick phases become indistinguishable \cite{schick_prew}. First-order wetting transitions have been
extensively studied both theoretically and experimentally.

\begin{figure*}
\centerline{\includegraphics[width=\linewidth]{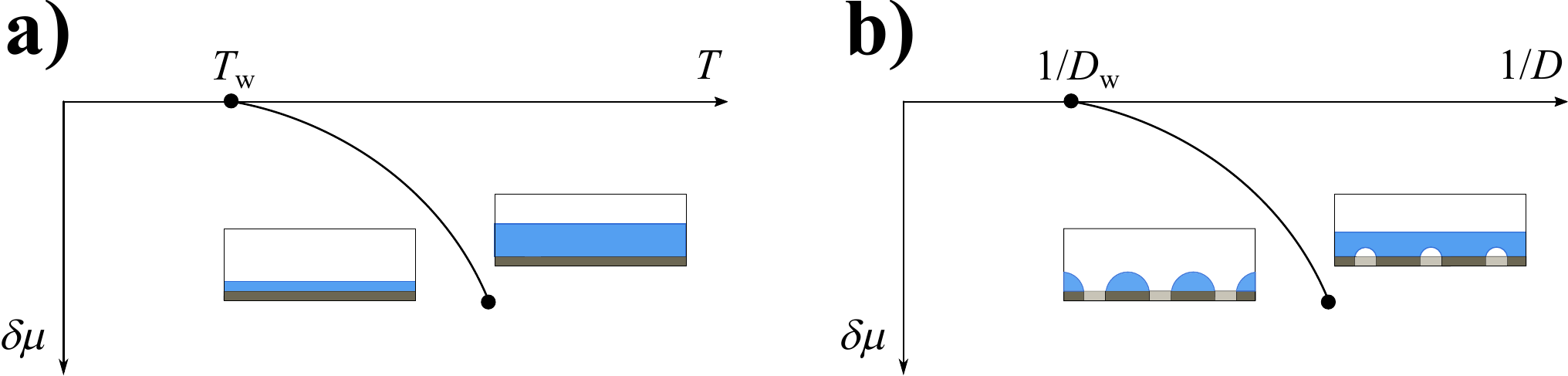}}
 \caption{Schematic surface phase diagrams representing first-order wetting transition a) on a
homogenous wall at a temperature $T_w$ and b) on a periodically nano-patterned wall of wet and dry stripes occurring when the distance $D$ between
the wet stripes is reduced to a value $D_w$. Also shown are the pre-wetting lines extending tangentially from the wetting transition off coexistence
$\delta\mu>0$ which terminate at a pre-wetting critical point. A schematic illustration of the phases which coexist along the pre-wetting line is
also shown.}\label{fig1}
\end{figure*}

When the wall is geometrically sculpted or chemically patterned the fluid adsorption can exhibit a zoo of new possibilities \cite{napior, parry2000,
bauer, groove_parry, cpwf, mal_groove, groove_mal, carlos, depinning, tewes}. Attempts to understand this date back to the empirical work of Wenzel
\cite{wenzel36} (for sculpted or rough surfaces) and Cassie \cite{cassie} for chemically patterned (planar) walls. For example, Cassie's ``law"
states that the effective contact angle $\theta_{\rm eff}$ on a composite substrate formed of two different materials, with different contact angles
$\theta_1$ and $\theta_2$, satisfies $\cos\theta_{\rm eff}=f\cos\theta_1+(1-f)\cos\theta_2$ where $f$ and $(1-f)$ are the fractions of the respective
wall areas of the different materials. However, more recent studies have shown that the situation is considerably richer due to new phase
transitions, absent in this simple empirical picture, which are induced by the nano-patterning of the substrate. For example, when two semi-infinite
materials meet at a line, a pre-wetting transition on either side can induce the lateral growth of a thick wetting layer which continuously spreads
out across the surface \cite{com_prew}. Transitions also occur when a partially wet surface ($\theta>0$) is decorated with stripes of a material
which is completely wet ($\theta=0$). When the distance $D$ between the stripes is large separate liquid drops nucleate above each stripe. Analysis
based on mesoscopic, interfacial Hamiltonian, models  predict that for systems with long-ranged dispersion forces the maximum (mid-point) height of
the drop scales, at bulk coexistence, as $h_m\propto\sqrt{L}$ where $L$ is the strip width \cite{mpp}. However as the inter-strip distance $D$ is
reduced a single (larger) drop forms which spans both stripes and the area between them. For systems with long-ranged forces this bridging transition
is predicted to occur when $D\propto\ln L$ which is always a molecularly small distance. The logarithmic dependence here directly arises from the
finite-size contribution to the surface free-energy of a drop above a completely wet stripe.

There is in fact a close connection between bridging and wetting transitions \cite{three_stripes}. Consider, for example, a generalization of the
above scenario where $N$ completely wet stripes decorate a partially wet substrate in contact with a gas at bulk coexistence. Again, when the
distance $D$ between the stripes is large, each nucleates a separate liquid drop above it. As $D$ is reduced, to a value close to $\ln L$, there is a
first-order phase transition to a single droplet which spans all the stripes and intervening (partially wet) spaces. For systems with dispersion
forces the height of this single large drop scales as $h_m\propto \sqrt{NL}$. It follows that in the limit $N\to\infty$, corresponding to a periodic
array of stripes, the bridging transition is equivalent to the unbinding of a liquid-gas interface from the wall i.e. a first-order wetting
transition. There are several aspects of this structurally induced wetting transition which are of interest. Firstly, it is driven by microscopic
forces and is not captured by the approximate Cassie equation which predicts that $\theta_{\rm eff}=0$ only occurs when $f=1$ (i.e. when the distance
between the stripes regions vanishes). Secondly, if there is a first-order wetting transition when $D=D_w\propto\ln L$ then there should be a line of
transitions analogous to pre-wetting occurring off coexistence for $D<D_w$ where a phase consisting of isolated drops coexist with a liquid film of
finite thickness which covers the whole substrate. We illustrate this in Fig.~\ref{fig1}b for the extreme scenario when the regions between the
stripes are completely dry ($\theta=\pi$). In this case the low adsorption phase consists of isolated liquid drops covering the wet stripes while the
high adsorption phase has bubbles over the dry regions and an additional liquid-gas interface separating the liquid slab from the bulk gas. Now, if
there is a pre-wetting line then we should expect that it terminates at pre-wetting critical point where the coexisting low and high adsorption
phases are indistinguishable. However, if this is the case we are left with the question as to how the two phases as sketched in Fig.~\ref{fig1}
become indistinguishable since they appear to be structurally different. To understand this we must abandon mesoscopic approaches which rely on
simple pictures of interfacial configurations and apply fully microscopic theory based on molecular density profiles. The purpose of the present
paper is study such bridging induced wetting and pre-wetting transitions using microscopic Density Functional Theory (DFT). In addition to
demonstrating that these transitions take place we also show, by varying the stripe width $L$, that the locations of the pre-wetting lines show
simple scaling data collapse associated with the presence of dispersion forces.

\section{Density functional model}

Within the framework of classical DFT \cite{evans79}, the equilibrium density profile $\rho_{\rm eq}(\rr)$, is obtained from the minimization of a grand
potential functional $\Omega[\rho]$. This is exactly written
 \bb
 \Omega[\rho]={\cal F}[\rho]+\int\dd\rr\rhor[V(\rr)-\mu]\,,\label{om}
 \ee
where $V(\rr)$ is the external potential modelling the wall and ${\cal F}[\rho]$ is the intrinsic Helmholtz free-energy functional which contains all
the information about the fluid-fluid intermolecular interaction. The formulation of DFT is exact in principle and higher derivatives of the
free-energy functional determine direct correlation functions and in turn pair-wise density-density correlations from solution of the inhomogeneous
Ornstein--Zernike equation. In most applications of DFT, approximations for ${\cal F}[\rho]$ must be made. Varieties of these have been developed
extensively over the last few decades to accurately model short-ranged intermolecular repulsive forces and long-ranged intermolecular attractions. It
is standard to separate the Helmholtz free energy  into an exact ideal gas contribution and an excess part:
  \bb
  {\cal F}[\rho]=\beta^{-1}\int\dr\rho(\rr)\left[\ln(\rhor\Lambda^3)-1\right]+{\cal F}_{\rm ex}[\rho]\,,
  \ee
where $\Lambda$ is the thermal de Broglie wavelength and $\beta=1/k_BT$ is the inverse temperature. Most modern DFT approaches follow the spirit of
van der Waals and the excess part is modelled as a sum of hard-sphere and attractive contributions where the latter is treated in simple mean-field
fashion:
  \bb
  {\cal F}_{\rm ex}[\rho]={\cal F}_{\rm hs}[\rho]+\frac{1}{2}\int\int\dd\rr\dd\rr'\rhor\rho(\rr')u_a(|\rr-\rr'|)\,, \label{f}
  \ee
where  $u_a(r)$ is the attractive part of the fluid-fluid interaction potential.

The fluid atoms are assumed to interact with one another via the truncated (i.e., short-ranged) and non-shifted Lennard-Jones-like potential
 \bb
 u_a(r)=\left\{\begin{array}{cc}
 0\,;&r<\sigma\,,\\
-4\varepsilon\left(\frac{\sigma}{r}\right)^6\,;& \sigma<r<r_c\,,\\
0\,;&r>r_c\,.
\end{array}\right.\label{ua}
 \ee
which is cut-off at $r_c=2.5\,\sigma$, where $\sigma$ is the hard-sphere diameter.

The hard-sphere part of the excess free energy is approximated using the fundamental measure theory (FMT) functional \cite{ros},
 \bb
{\cal F}_{\rm hs}[\rho]=\frac{1}{\beta}\int\dd\rr\,\Phi(\{n_\alpha\})\,,\label{fmt}
 \ee
which accurately takes into account the short-range correlations between fluid particles allowing for an accurate description of layering arising
from the volume exclusion when a high density liquid is in contact with a wall. We have adopted the original Rosenfeld theory where there are six
weighted densities $n_\alpha$ which are themselves convolutions of the density profile and fundamental measures of the hard sphere of diameter
$\sigma$.

The external potential $V(\rr)$ is chosen to model a periodic array of stripes of width $L$ separated by regions of width $D$. Here $\rr =(x,z)$
where $x$ and $z$ are the coordinates along and perpendicular to the wall, respectively. Translational invariance is assumed along the stripes and
the potential is infinite for $z<0$ i.e. a hard-wall repulsion. However in the stripe regions we assume that there is a strong attraction between the
fluid particles and the substrate atoms which are assumed to be uniformly distributed with density $\rho_w$. The potential $V(\rr)$ for $z>0$ is then
determined by integrating $\rho_w\phi_w(r)$ over the volume of whole array of stripes. Here $\phi_w(r)$ is the long-ranged attractive tail of the
Lennard-Jones potential
 \bb
 \phi_w(r)=-4\varepsilon_w\left(\frac{\sigma}{r}\right)^{6}\,,
 \ee
which has a strength parameter $\varepsilon_w$ chosen to emulate complete wetting if the striped region covered the whole surface. Hence, the
attractive part of wall potential can be written as
 \begin{eqnarray}
  V(x,z)=\sum_{n=-\infty}^{\infty} V_{L}(x+nL,z)\,. \label{v1}
 \end{eqnarray}
where
 \begin{eqnarray}
V_{L}(x,z) &=&\alpha_w\left[\frac{1}{(x-L)^3}-\frac{1}{x^3}\nonumber\right.+\\
            &&\left.+\psi_6(x-L,z)-\psi_6(x,z)\right]
\end{eqnarray}
with
 \bb
\alpha_w=-\frac{1}{3}\pi\varepsilon_w\sigma^6\rho_w
 \ee
 and
 \bb
\psi_6(x,z)=-{\frac {2\,{x}^{4}+{x}^{2}{z}^{2}+2\,{z}^{4}}{2{z}^{3}{x}^{3} \sqrt {{x}^{2}+{z}^{2}}}}\,.
 \ee

 \begin{figure}
\centerline{\includegraphics[width=\linewidth]{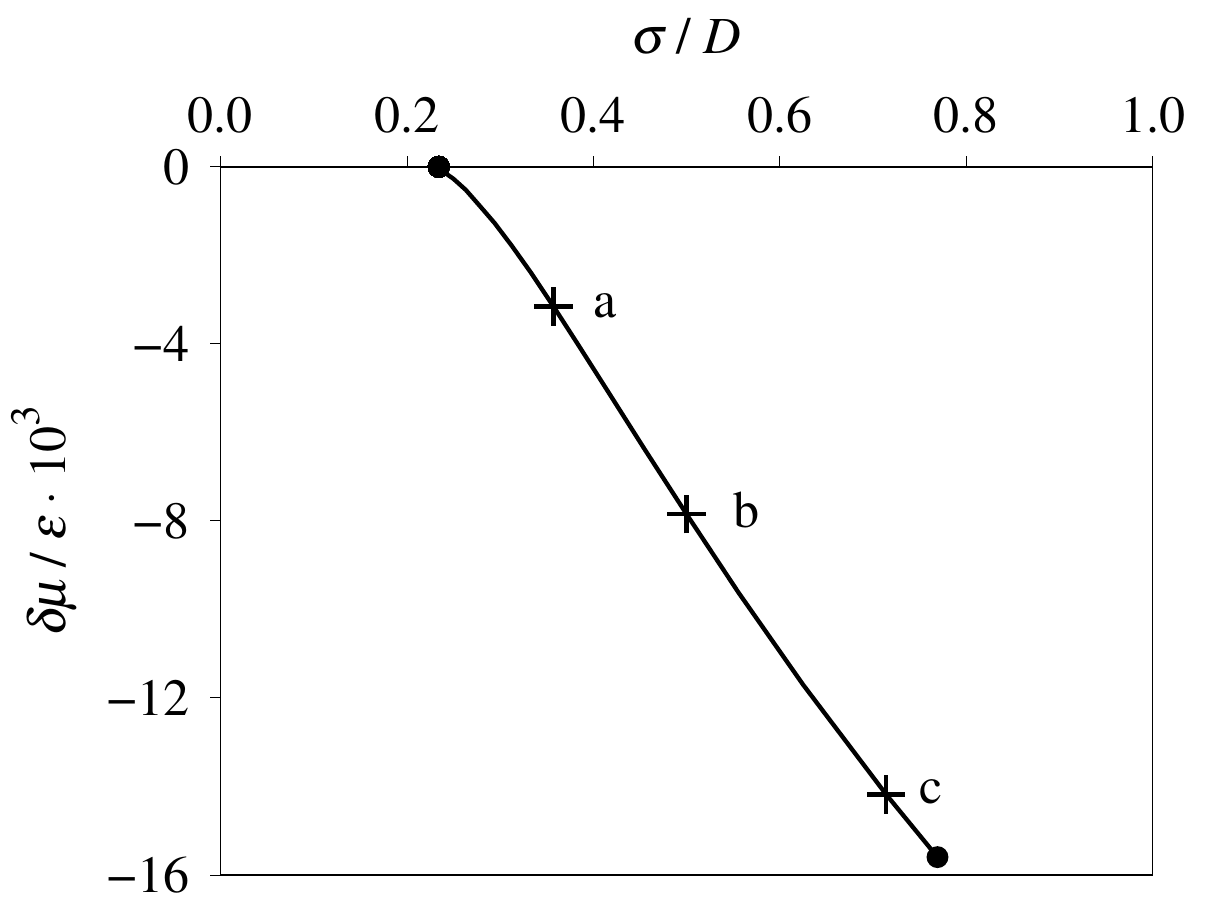}} \caption{Numerical DFT results showing the locations of the wetting and pre-wetting transitions
on a dry hard wall patterned with completely wet stripes of width $L=30\,\sigma$.}\label{fig2}
\end{figure}

In our substrate model the region in between the stripes is modelled by a simple hard-wall potential meaning that these parts of the substrate would
be completely dry at any temperature if they were of macroscopic extend. In our calculations we have chosen $\varepsilon_w=\varepsilon$ for which the
wetting temperature of the homogenous wall $T_w=0.8\,T_c$ where $T_c$ is the bulk critical temperature. By choosing $T=0.92\,T_c$ we therefore ensure
that the stripes are completely wet by liquid.

Minimization of (\ref{om}) leads to the Euler-Lagrange equation
 \bb
 V(\rr)+\frac{\delta{\cal F}_{\rm hs}}{\delta\rho(\rr)}+\int\dd\rr'\rho(\rr')u_{\rm a}(|\rr-\rr'|)=\mu\,,\label{el}
 \ee
which can be solved iteratively on an appropriately discretized two dimensional grid $(0,x_m)\times(0,z_m)$ where $x_m=L+D$ represents one period and
$z_m=50\,\sigma$. Periodic boundary conditions are assumed at the edges in the $x$ direction and the density far from the wall at $z_m=50\,\sigma$ is
fixed to the equilibrium bulk vapour.

\section{Results: scaling of the pre-wetting lines}

\begin{figure*}
\centerline{\includegraphics[width=\linewidth]{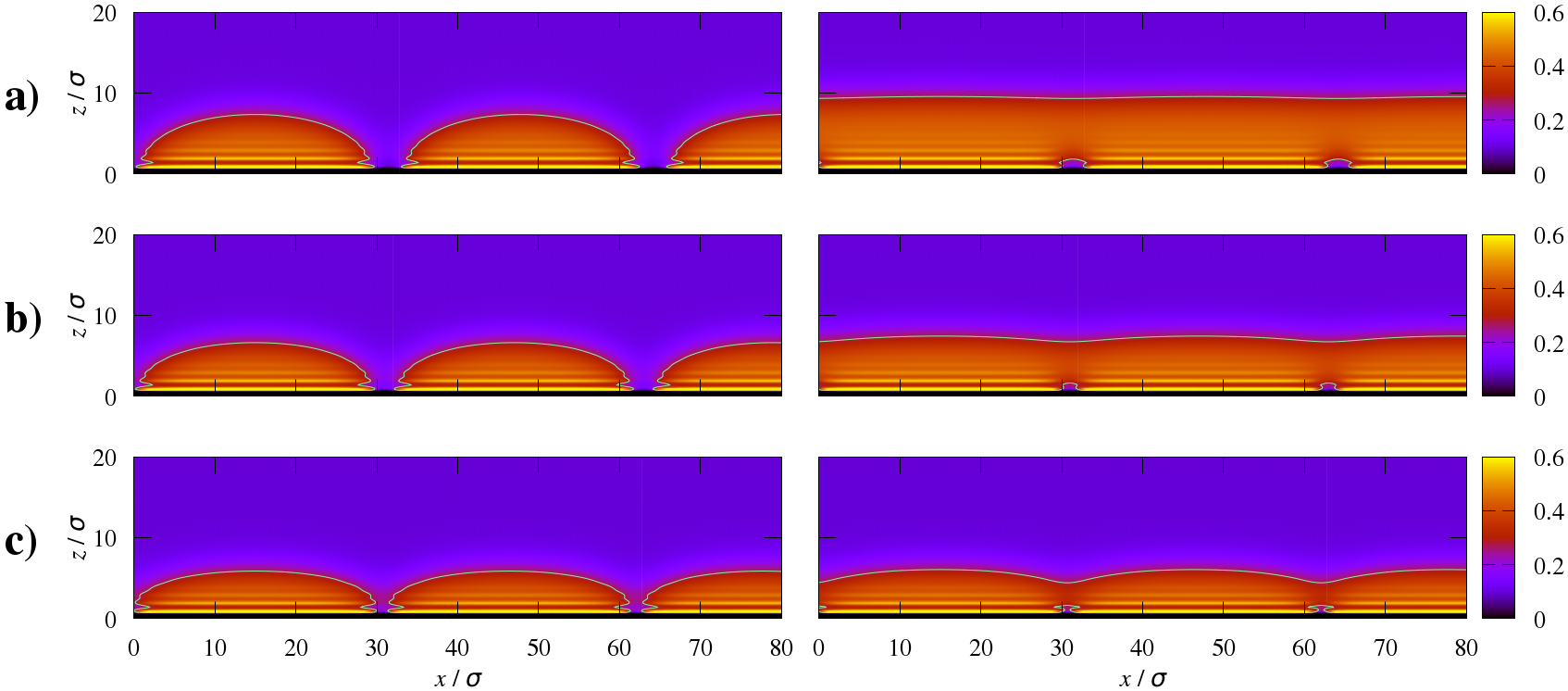}} \caption{Coexisting density profiles at three representative points along the pre-wetting for
$L=30\,\sigma$ as highlighted in Fig~\ref{fig2}. Close to bulk coexistence the coexisting phases are structurally different and droplets and bubbles
over the wet and dry stripes can be distinguished. However, on approaching the pre-wetting critical point the interfacial structure very close to the
wall near the dry regions becomes more diffuse and the phases become increasingly similar.}\label{fig3}
\end{figure*}

\begin{figure}
\centerline{\includegraphics[width=\linewidth]{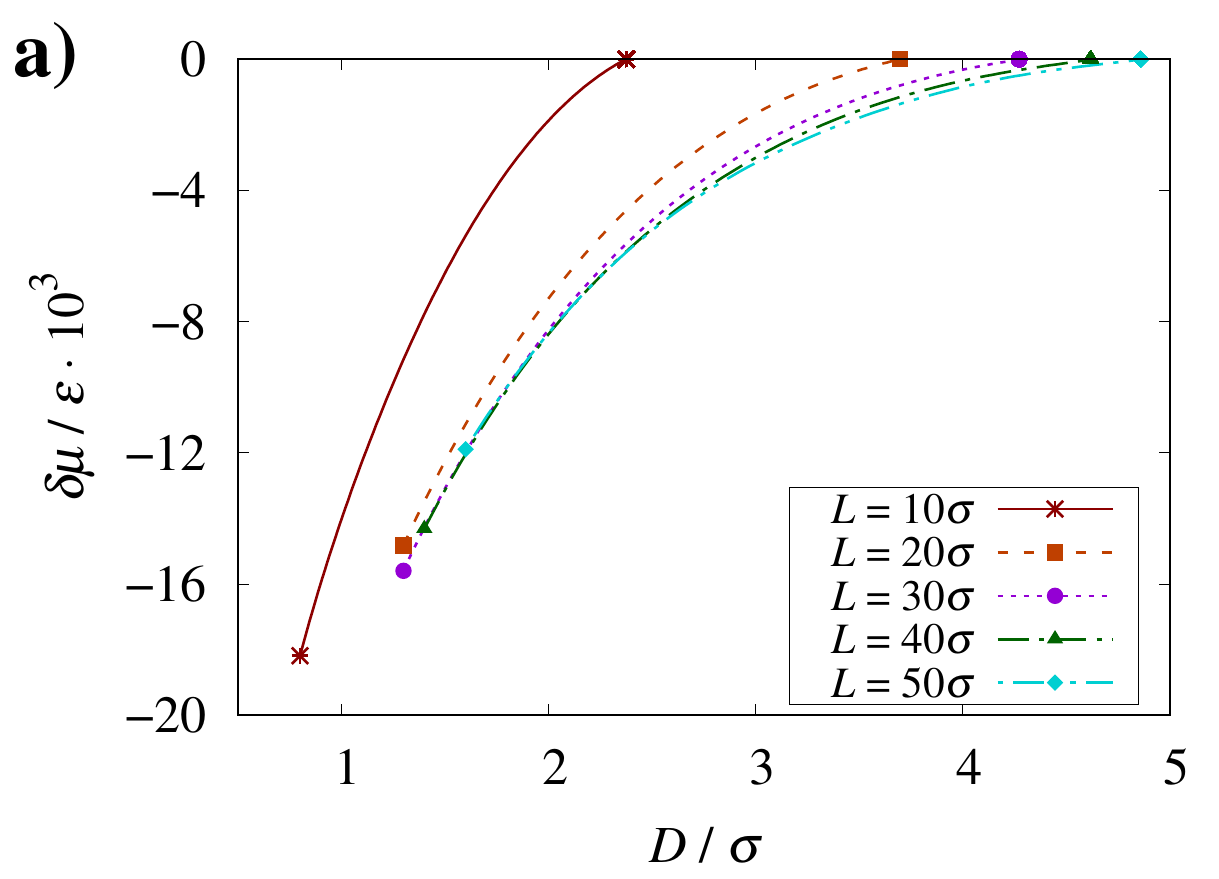}} \centerline{\includegraphics[width=\linewidth]{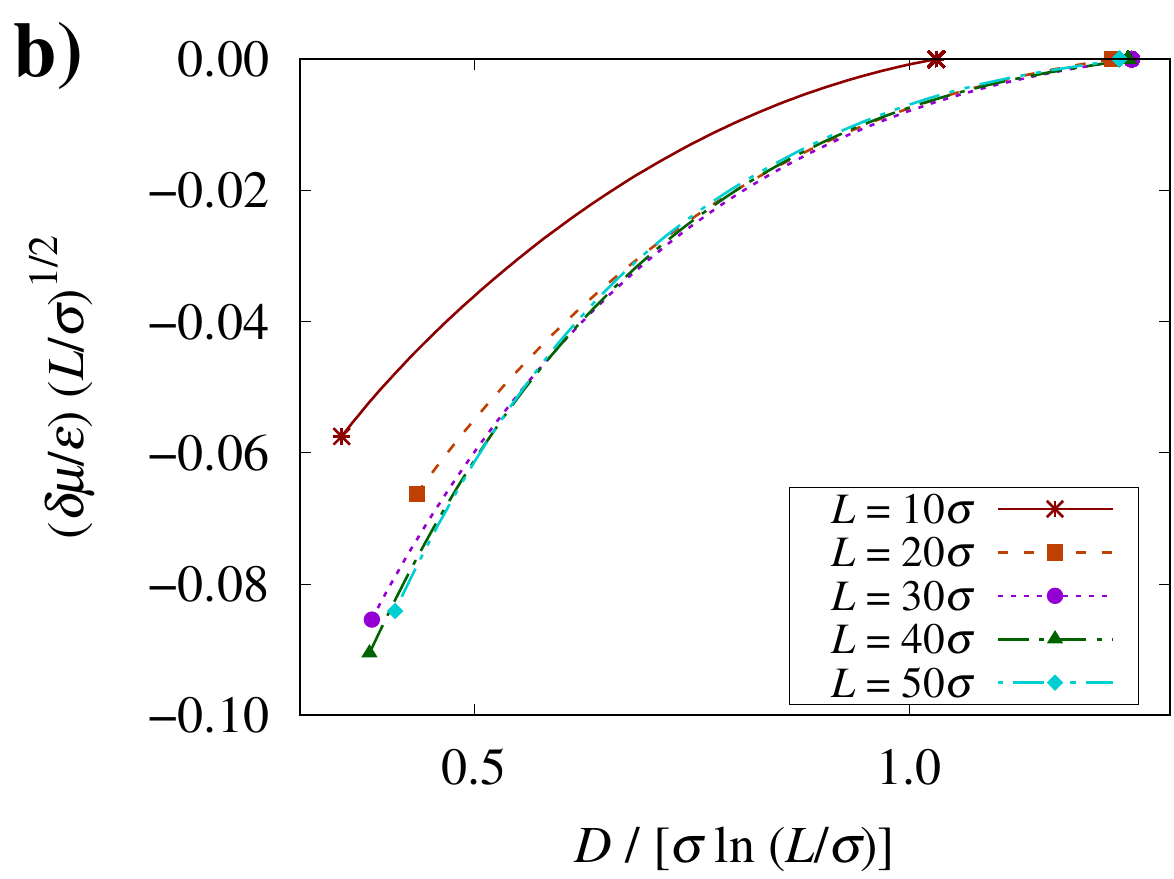}} \caption{Numerical DFT results showing
a) the wetting and pre-wetting lines for five different stripe widths and b) the data collapse when the separation distance $D$ and chemical
potential field $\delta\mu$ are rescaled by $\ln L/\sigma$ and $\sqrt{L/\sigma}$, respectively. }\label{fig4}
\end{figure}

Fig.~\ref{fig2} shows the numerically determined phase diagram for a patterned wall with stripes of width $L=30\,\sigma$. As anticipated, at bulk
coexistence, the system shows a first order wetting transition at $D_w= 4.27\sigma$ at which a low adsorption phase consisting of isolated drops
coexists with a completely wet state where a macroscopic layer of liquid covers the surface. The value $D_w$ is determined by balancing the
grand-potentials of these configurations, equivalent to an application of Antonoff's rule $\gamma_{wg}=\gamma_{wl}+\gamma$ which determines that the
contact angle on the substrate $\theta_{\rm eff}=0$. Here $\gamma_{wg}$ is the surface tension of the wall-gas interface (with droplets of liquid
over the stripes) $\gamma_{wl}$ is the surface tension of the wall-liquid interface (with bubbles of gas over the dry interstitial gaps) and $\gamma$
is the usual liquid-gas surface tension of a planar interface. A similar matching of grand potentials determines the pre-wetting line extending away
tangentially in the $1/D$-$\delta\mu$ plane. Representative coexisting density are shown in Fig.~\ref{fig3}. Close to bulk coexistence we can easily
distinguish drop and bubble like structures which cover the wet or dry patches respectively (see Fig.~\ref{fig3}a). However on approaching the
pre-wetting critical point, the value of the stripe separation $D$ is so small that one can no longer detect bubble configurations over the dry
patches so that the density close to the wall is similar for both thin and thick film phases. That is they become more and more structurally similar.
At the pre-wetting critical point the total adsorption $\Gamma=\int dxdz(\rho(\rr)-\rho_g)$ is the same for both phases.

Finally, in Fig.~\ref{fig4}a we show the phase diagrams obtained for five different stripe widths $L=10\,\sigma,20\,\sigma,30\,\sigma,40\,\sigma$,
and $50\,\sigma$ expressed simply in the $D-\delta\mu$ plane. All the pre-wetting lines have a similar shape and it is apparent that as $L$ increases
the value of $D_w$ increases. In fact we should anticipate that there is some data collapse which encapsulates universal features. Recall that for
dispersion forces the Clausius-Clapeyron equation determines that, close to bulk co-existence, the pre-wetting line $T_{pw}(\delta\mu)$ behaves as
$T_w-T_{pw}\propto(\delta\mu)^{2/3}$ \cite{dietrich}. Here the power-law dependence reflects the singular contribution to the excess surface free
energy associated with the complete wetting transition. The constant of proportionality missing here is non-universal, system dependent, and related
to the value of the Hamaker constant. A similar result applies to the present heterogeneous system and all the pre-wetting lines $D_{pw}(\delta\mu)$
behave, close to bulk coexistence, as $D_{pw}-D_w\propto(\delta\mu)^{2/3}$. In fact the only difference between pre-wetting curves shown in
Fig.~\ref{fig4}a is due to the stripe width $L$ -- otherwise the systems are chemically identical. It follows that if we rescale $D$ and $\delta\mu$
allowing for their dependence on the stripe width $L$, there should be some data collapse. Fig.~\ref{fig4}b shows the data collapse obtained when $D$
is rescaled with $\ln L/\sigma$ (as predicted for the value of $D_w$) and $\delta\mu$ by $L^{1/2}$ (representing the height of the droplets
equivalent to their volume per unit area). Apart from the very smallest system $L=10\sigma$ there is excellent data collapse for the shape of the
pre-wetting line particularly close to bulk coexistence.  This  therefore verifies the  predictions that the wetting transition occurs at $D_w\propto
\ln L$ and that the droplet height scales as $h_m\propto\sqrt{L}$. This is the main finding of our study.

\section{Concluding remarks}

 Our study has been based on a microscopic but mean-field DFT model which misses some fluctuation effects. However, unlike the bridging transition
between two (or any finite number) of stripes, which beyond mean-field would be rounded by fluctuations, the wetting and pre-wetting transitions
considered here remain genuine first-order as predicted by mean-field theory. The present analysis should therefore accurately predict the value of
the stripe spacing $D_w$ where the wetting transition occurs. Secondly the scaling dependence seen in the data collapse for the different pre-wetting
lines is unchanged by interfacial fluctuations since they arise from critical singularities associated with complete wetting, the upper critical
dimension for which is less than three for dispersion forces \cite{forgacs}. The only inaccuracy in the DFT analysis concerns predictions for the
pre-wetting critical point which, beyond mean-field, should belong to the true 2D Ising bulk universality class \cite{nakanishi} . An interesting
extension of the present work would be to consider arrays of stripes with different widths which could lead to additional bridging transitions and
the sequential coalescence of drops which precede a wetting transition. It would also be interesting to consider the effect of disorder, by
randomizing the stripe widths, which would more realistically model experimental situations. Finally, it might be worth investigating the effect of
the model fluid potential; e.g. by considering a long-ranged fluid-fluid potential we expect much stronger interaction between the neighboring
droplets facilitating their coalescence.

In summary, we have used a microscopic DFT to demonstrate that  bridging induced wetting transitions occur on nano-patterned walls comprising wet and
dry stripes. The associated pre-wetting lines occurring in systems with different stripe widths show a simple data collapse which confirms scaling
predictions for the height and finite-size surface free energy for complete wetting drops.

\begin{acknowledgments}
\noindent This work was funded in part by the EPSRC UK Grant no. EP/L020564/1, ``Multiscale Analysis of Complex Interfacial Phenomena". A.M.
acknowledges the support from the Czech Science Foundation, Project No. GA17-25100S. M.P. acknowledges the financial support from specific university
research (MSMT No 21-SVV/2019).
\end{acknowledgments}

\end{document}